\documentclass[aps,prc,twocolumn,superscriptaddress,showkeys,nofootinbib]{revtex4-1}  

\usepackage[utf8]{inputenc}

\usepackage{graphicx,color}
\usepackage{amsmath,amssymb,amsfonts}
\usepackage{hyperref}

\usepackage{xspace}	% Include xspace

\newcommand{\pcct}{\mbox{$\rho(v^{2}_{2},\langle p_{T} \rangle)$}\xspace}

\newcommand{\roots} {\mbox{$\sqrt{\textit{s}_{NN}}$}\xspace}

\def  \fourth    {\mbox{$\textit{v}_{4}$}\xspace}

\def  \etas      {\mbox{$\eta / \textit{s}$ }\xspace}

\def \sc23  {\mbox{$\mathrm{SC}(2,3)$   }\xspace}
\def \sc24  {\mbox{$\mathrm{SC}(2,4)$   }\xspace}

\def \nsc23 {\mbox{$\mathrm{NSC}(2,3)$}\xspace}
\def \nsc24 {\mbox{$\mathrm{NSC}(2,4)$}\xspace}

\begin{document}
\title{Impact of nuclear deformation on collective flow observables in relativistic U+U collisions}
%Systematic study of the nuclear quadrupole deformation in U$+$U collisions at RHIC
%------------------------------------------------------------------------
\medskip
%------------------------------------------------------------------------
\author{Niseem~Magdy} 
\email{niseemm@gmail.com}
\affiliation{Department of Chemistry, State University of New York, Stony Brook, New York 11794, USA}
%------------------------------------------------------------------------
%------------------------------------------------------------------------
%\date{\today}
%------------------------------------------------------------------------
\begin{abstract}
%------------------------------------------------------------------------
%------------------------------------------------------------------------
A Multi-Phase Transport (AMPT) model is used to investigate the efficacy of several flow observables to constrain the initial-state deformation of the Uranium nuclei in U$+$U collisions at nucleon-nucleon center-of-mass energy \roots = 193 GeV. 
The multiparticle azimuthal cumulant method is used to investigate the sensitivity of (I) a set of quantities that are sensitive to both initial- and final-state effects as well as (II) a set of dimensionless quantities that are more sensitive to initial-state effects to the Uranium nuclei quadrupole shape deformation.
We find that the combined use of the flow harmonics, flow fluctuations and correlations, linear and non-linear flow correlations to the quadrangular flow harmonic, and the correlations between elliptic flow and the mean-transverse momentum could serve to constrain the nuclear deformation of the Uranium nuclei.
Therefore, a comprehensive set of measurements of such observables can provide a quantifying tool for the quadrupole shape deformation via data-model comparisons.
%------------------------------------------------------------------------
%------------------------------------------------------------------------
\end{abstract}
%------------------------------------------------------------------------
%\pacs{25.75.-q, 25.75.Gz, 25.75.Ld}% PACS, the Physics and Astronomy
                             % Classification Scheme.
%------------------------------------------------------------------------
%------------------------------------------------------------------------
\maketitle
%------------------------------------------------------------------------
%\linenumbers
%------------------------------------------------------------------------

\section{Introduction}
%------------------------------------------------------------------------
%Introduction
%------------------------------------------------------------------------
%------------------------------------------------------------------------
The quark-gluon plasma (QGP) -- a new state of matter, is produced in ultra-relativistic heavy-ion collisions~\cite{Shuryak:1978ij,Shuryak:1980tp,Muller:2012zq}. 
{\color{black}Understanding the QGP's specific shear viscosity $\etas(T, \mu_{B})$ dependence on temperature ($T$) and baryon chemical potential ($\mu_{B}$)  is being investigated at the Relativistic  Heavy-Ion Collider (RHIC) and the Large Hadron Collider (LHC).}
%A key purpose of the heavy-ion programs at the Relativistic  Heavy-Ion Collider (RHIC) and the Large Hadron Collider (LHC) is to study the temperature ($T$) and baryon chemical potential ($\mu_{B}$) dependence of the QGP's specific shear viscosity $\etas(T, \mu_{B})$. 
{\color{black} This transport property describes the QGP's power to transfer initial-state anisotropies due to collision geometry into final-state momentum anisotropies.}
 Anisotropic flow ($v_n$) measurements have played a major role in elucidating this final-state effect since they {\color{black}originate from}  the viscous hydrodynamic response to the eccentricity ($\varepsilon_n$) of the energy-density distribution produced in the initial stages of the collision~\cite{Danielewicz:1998vz,Ackermann:2000tr,Adcox:2002ms,Heinz:2001xi,Huovinen:2001cy,Hirano:2002ds,Shuryak:2003xe,Hirano:2005xf,Romatschke:2007mq,Luzum:2008cw,Bozek:2009dw,Song:2010mg,Qian:2016fpi,Schenke:2011tv,Teaney:2012ke,Gardim:2012yp,Lacey:2013eia,Magdy:2020gxf,STAR:2002pmf,STAR:2001ksn,STAR:2000ekf,PHOBOS:2008yxa,PHENIX:2018lfu,PHENIX:2018wex,PHENIX:2002hqx,PHENIX:2014uik,PHOBOS:2010ekr,PHOBOS:2007wlf,PHOBOS:2006dbo,PHOBOS:2005ylx,CMS:2019nct,Magdy:2021sba,ALICE:2021adw,CMS:2019cyz,Magdy:2022jai,ATLAS:2017hap,ATLAS:2018ngv,ATLAS:2017rij}.
%------------------------------------------------------------------------

%------------------------------------------------------------------------
A possible deformation of the colliding nuclei can also influence $\varepsilon_n$ and consequently, $v_n$~\cite{Jia:2021tzt,Giacalone:2021udy}; the initial-state profile for each of the colliding nuclei can be characterized with the woods-Saxon distribution for the nuclear density as:
\begin{eqnarray}\label{eq:deformed_ws}
\rho & = & \frac{\rho_0}{1 + \exp{([r - R^{'}]/a)}} \nonumber,\\ 
R^{'}(\theta,\phi) & = & R [1 + \beta_2 Y_2^0(\theta,\phi) \nonumber\\
&+& \beta_3 Y_3^0(\theta,\phi) + \beta_4 Y_4^0(\theta,\phi) + ... ],
\end{eqnarray}
where $\rho_0$ is the {\color{black}nucleon density in the center of the nucleus}, $a$ is the {\color{black} skin depth} (\textit{i.e.}, the surface thickness), $R$ is the nuclear radius {\color{black}and is taken to be $1.2$ $A^{1/3}$~\cite{Shou_2015}}, and $R^{'}(\theta,\phi)$ is the nuclear surface which contains the relevant axial symmetric quadruple ($\beta_2$), octuple ($\beta_3$) and hexadecapole ($\beta_4$) deformations~\cite{Filip:2007tj,Bally:2021qys,Giacalone:2021udy,Rybczynski:2012av,Zhang:2022fou,Jia:2021oyt,Jia:2021tzt,Zhang:2021kxj,Jia:2021wbq}.
%------------------------------------------------------------------------
%
%
%,Zhao:2022uh
%------------------------------------------------------------------------
Within the quantum many-body system, the deformation is a fundamental property of the atomic nucleus that provides the correlated nature of the dynamics of nucleons. Many atomic nuclei exhibit a quadrupole or octupole deformation, which could influence the flow coefficients' magnitude, fluctuations, and correlations. Indeed, recent measurements employing detailed comparisons between Au$+$Au and U$+$U collisions~\cite{Adamczyk:2015obl} as well as Pb$+$Pb and Xe$+$Xe collisions~\cite{ALICE:2022xhd,ALICE:2018lao} have indicated signatures compatible with nuclear deformation. However, the degree to which flow measurements can provide constraints useful for detailed characterization of the deformation of colliding nuclei is still unclear.

%------------------------------------------------------------------------
Prior works have shown that the elliptic and triangular flow coefficients $v_{2}$ and $v_{3}$, are linearly related to the initial-state eccentricities, $\varepsilon_{2}$ and $\varepsilon_{3}$~\cite{Song:2010mg,Niemi:2012aj,Gardim:2014tya,Fu:2015wba,Holopainen:2010gz,Qin:2010pf,Qiu:2011iv,Gale:2012rq,Liu:2018hjh,Adam:2019woz}. The higher-order flow harmonics $v_{n>3}$ not only arise from a linear response to the same-order initial-state anisotropies, but also from a non-linear response to the lower-order eccentricities $\varepsilon_{2}$ and $\varepsilon_{3}$~\cite{Teaney:2012ke,Bhalerao:2014xra,Yan:2015jma}. The non-linear contributions  encode the correlations between different symmetry planes $\Psi_{n}$~\cite{ATLAS:2014ndd} which provides a constraint for the initial-stage dynamics~\cite{Bilandzic:2013kga,Bhalerao:2014xra,Aad:2015lwa,ALICE:2016kpq,STAR:2018fpo,Zhou:2016eiz,Qiu:2012uy,Teaney:2013dta,Niemi:2015qia,Zhou:2015eya}. The correlation between the flow harmonics $v_{n}$ and the event mean transverse momentum $\langle p_{T} \rangle$:
\begin{eqnarray}
    \rho(v^{2}_{n},\langle p_{T} \rangle) = \frac{{\rm cov}(v_{n}^{2},\langle p_{T} \rangle)}{\sqrt{{\rm Var}(v_{n}^{2})} \sqrt{{\rm Var}(\langle p_{T} \rangle)}}\label{eq:1},
\end{eqnarray}
which indicates different sensitivities to the initial- and final-states of the collisions~\cite{Bozek:2016yoj,Giacalone:2017uqx,Giacalone:2020byk,Bozek:2016yoj,Bozek:2020drh,Schenke:2020uqq,Giacalone:2020dln,Lim:2021auv,ATLAS:2022dov,Magdy:2021ocp} has also been shown to be sensitive to nuclear deformation, albeit with added sensitivity to the $p_T$ selection, event shape selection, $\eta$ selection, and the centrality definition~\cite{ATLAS:2022dov,ALICE:2021gxt}.
%------------------------------------------------------------------------
Here, we use detailed simulations with the AMPT model to investigate supplemental measures that are sensitive to initial-state deformation which could be used to constrain nuclear deformation.

In the current work, we investigate the influence of the the nuclear quadrupole deformation ($\beta_{2} > 0.0$) in U$+$U collisions at nucleon-nucleon center-of-mass energy \roots = 193~GeV on the $v_{n}\lbrace k\rbrace$~\cite{Giacalone:2018apa,Jia:2021tzt,Giacalone:2021udy}, $v_{2}\lbrace 2\rbrace$/$v_{2}\lbrace 4\rbrace$, 
the normalized symmetric cumulants ($\rm NSC(2,3)$),
the linear and non-linear contributions to the $\textit{v}_{4}$, the coupling constant ($\chi_{4,22}$), the correlations between different order flow symmetry planes ($\rho_{4,22}$), and the correlation between the flow harmonics and the event mean $p_T$, $\rho(v^{2}_{n},\langle p_{T} \rangle)$~\cite{Jia:2021wbq,Bally:2021qys,Giacalone:2021clp,Liyanage:2022nua}.
Here, an important objective is to develop a more stringent  constraint for initial-state deformation by simultaneously leveraging the response of of several correlators to nuclear quadrupole deformation of the Uranium nuclei. 
%
% 
%The current sensitivity study, conducted within the AMPT model~\cite{Lin:2004en} framework could lend important insights on how to constrain the initial-state models. This will lead to a better description of the system size dependence of the presented variables~\cite{STAR:2019zaf,Magdy:2020bij,STAR:2022gki}.
%------------------------------------------------------------------------

%------------------------------------------------------------------------
The paper is organized as follows.
Section~\ref{sec:2} summarizes the theoretical model used to investigate the  $\beta_2$ dependence on the flow quantities and the details of the analysis method employed. The results from the model studies are presented in Sec.~\ref{sec:3} followed by a summary in Sec.~\ref{sec:4}.

%------------------------------------------------------------------------
%------------------------------------------------------------------------
\section{Methodology} \label{sec:2}
%------------------------------------------------------------------------
%The models used in this work are described in \ref{sec:2a} and the analysis technique used to compute  $\rho(v^{2}_{2},\langle p_{T} \rangle)$ is given in \ref{sec:2b}.

\subsection{The AMPT model}\label{sec:2a}
%------------------------------------------------------------------------
{\color{black}The current study is performed with simulated events for U+U collisions at \roots = 193~GeV, generated using the AMPT~\cite{Lin:2004en} model.  Analysis was performed for charged hadrons in the transverse momentum range of $0.2$ $<$ $p_T$ $<$ $2.0$~GeV/c and the pseudorapidity acceptance $|\eta|$ $<$ $1.0$. In addition, the simulated events were partitioned into several collision centrality classes established on the collision's impact parameter.}

%------------------------------------------------------------------------
The AMPT  model~\cite{Lin:2004en} is widely used to study the physics of the relativistic heavy-ion collisions at LHC and RHIC~\cite{Lin:2004en,Ma:2016fve,Haque:2019vgi,Bhaduri:2010wi,Nasim:2010hw,Xu:2010du,Magdy:2020bhd,Guo:2019joy}. 
In this work, events were generated with the AMPT model with the string melting option. In such a scenario {\color{black} initial conditions are given using the Glauber model}, hadrons are created using the HIJING model and converted to their valence quarks and anti-quarks, and their space-time evolution is evaluated via the ZPC Parton cascade model~\cite{Zhang:1997ej}.
The fundamental elements of the AMPT model are (i) the HIJING model~\cite{Wang:1991hta,Gyulassy:1994ew} initial Parton-production stage, (ii) a Parton-scattering stage, (iii) hadronization via coalescence  then (iv)  a hadronic interaction stage~\cite{Li:1995pra}.  
In the Parton-scattering stage the utilized Parton-scattering cross-sections are evaluated using;
\begin{eqnarray} \label{eq:21}
\sigma_{pp} &=& \dfrac{9 \pi \alpha^{2}_{s}}{2 \mu^{2}},
\end{eqnarray}
where $\alpha_{s}$ is the QCD coupling constant and $\mu$ is the screening mass in the partonic matter. They typically give the expansion dynamics of the A--A collision systems~\cite{Zhang:1997ej}; 
In this work, U$+$U collisions at $\sqrt{s_{\rm NN}}=$ 193~GeV, were simulated for a fixed value of $\alpha_{s}$ = 0.47, and $\mu$ = 3.41~$fm^{-1}$~\cite{Xu:2011fi,Nasim:2016rfv}.
%------------------------------------------------------------------------

%------------------------------------------------------------------------
The U+U collisions are implemented in the AMPT model by parametrizing the nucleon density distribution as a deformed Woods-Saxon profile~\cite{Hagino:2006fj} that is given in Eq.~\ref{eq:1}.
%------------------------------------------------------------------------
%\begin{eqnarray}\label{eq:deformed_ws}
%\rho & = & \frac{\rho_0}{1 + \exp{([r - R^{'}]/a)}} \nonumber,\\ 
%R^{'}(\theta,\phi) & = & R~ [1 + \beta_2 Y_2^0(\theta,\phi) \nonumber,\\
%&+& \beta_3 Y_3^0(\theta,\phi) + \beta_4 Y_4^0(\theta,\phi) + ... ].
%\end{eqnarray}
%------------------------------------------------------------------------
%------------------------------------------------------------------------
%The $4\pi r^2 \sin{(\theta)} \rho(r) ~ d\theta d\phi$ gives the  nucleons positions. 
In U$+$U collisions the projectile and target nuclei are rotated randomly event-by-event along the polar and azimuthal directions. 
%------------------------------------------------------------------------
%------------------------------------------------------------------------
The nucleon density distribution Eq.~\ref{eq:deformed_ws}, as well as the initial state eccentricities, can be varied by adjusting the values of; {\color{black} (i) The parameter $a$ that is generally used to distinguish the nucleon sampling from Woods-Saxon distribution between protons and neutrons, (ii) The $\beta_2$ which describes the overall quadrupole deformation, (iii) The $\beta_3$ that controls the overall octupole deformations, and (iv) The $\beta_4$ that describes the hexadecapole deformation. Prior investigations inspire the model parameters in this work~\cite{Shou_2015,DEVRIES1987495}. Note that the $\beta$ parameters influence the magnitude of the eccentricity (and its fluctuations) in central collisions.} In the current work $a$ is fixed to $a=0.44$, and the deformation parameters~\cite{Moller:1993ed} for Uranium are given in Tab.~\ref{tab:1}.

%------------------------------------------------------------------------
%------------------------------------------------------------------------
% Tab. 1an
%------------------------------------------------------------------------
\begin{table}[h!]
\begin{center}
 \begin{tabular}{|c|c|c|c|}
 \hline %Mechanism
 AMPT-set      &         $\beta_{2}$      &    $\beta_{4}$   \\
  \hline
  Set-1        &           0.00           &   0.000          \\
  \hline
  Set-2        &           0.00           &   0.200          \\
 \hline
  Set-3        &           0.28           &   0.093          \\
 \hline 
  Set-4        &           0.40           &   0.093          \\
 \hline
\end{tabular} 
\caption{The summary of the AMPT sets used in this work.}
\label{tab:1}
\end{center}
\end{table}
%------------------------------------------------------------------------%------------------------------------------------------------------------
\subsection{Analysis Method}\label{sec:2b}
%------------------------------------------------------------------------
%------------------------------------------------------------------------
The two- and multi-particle cumulants methods are used in this work.
{\color{black}The cumulants method was initially introduced in Refs~\cite{Borghini:2000sa,Borghini:2001vi} and was extended using (i) the Q-cumulants~\cite{Bilandzic:2010jr}, (ii) the Generic framework~\cite{Bilandzic:2013kga}, and (iii) the Subevent cumulants methods~\cite{Jia:2017hbm}.}
%--------------------------------------------------------------------------------------------------------------------------------------------
The two- and multi-particle cumulants can be constructed in terms of n$^{th}$ flow vectors ($Q_{n}$) magnitude. The  $Q_{n}$ are given as:
%------------------------------------------------------------------------
%{\footnotesize 
\begin{eqnarray}\label{eq:21-1}
Q_{n,k}              &=&  \sum^{M}_{i=1} \omega^{k}_{i} e^{in\varphi_{i}},
\end{eqnarray}
%}
%------------------------------------------------------------------------
where $M$ is the total number of particles in an event and  $\omega_{i}$ is the $\textit{i}^{th}$ particle weight, note that for uniform acceptance $\omega_{i} = 1$. Also  the sum over the particles weight is introduced as:
%------------------------------------------------------------------------
%{\footnotesize 
\begin{eqnarray}\label{eq:21-2}
S_{p,k}              &=&  \left[  \sum^{M}_{i=1} \omega^{k}_{i}  \right]^{p}.
\end{eqnarray}
%}

Using Eqs.(\ref{eq:21-1}, \ref{eq:21-2}) the two-, three-,  and four-particle correlations were constructed using the two-subevents cumulant methods~\cite{Jia:2017hbm},  with  $|\Delta\eta| = |\eta_{a} - \eta_{b}|  > 0.7$ ($\eta_{a}~ > 0.35$ and $\eta_{b}~ < -0.35$).
%------------------------------------------------------------------------
%\left(  v^{\rm Inclusive}_{n}  \right)^{2}     &=&   v^{2}_n\{2\} 
%{\small

\begin{itemize}

%------------------------------------------------------------------------
\item{The two-particle correlations:}
%------------------------------------------------------------------------
\begin{eqnarray}\label{eq:21-3}
v^{2}_n\{2\}  &=& \left\langle  \left\langle  2 \right\rangle \right\rangle_{n} , \nonumber \\
 &=& \langle\langle  \cos(n\phi^{a}_1 - n\phi^{b}_2 \rangle\rangle , \nonumber \\
&=&   \sum^{N_{ev}}_{i=1} (\mathcal{M}_{2})_{i} \left\langle  2 \right\rangle_{n,i} / \sum^{N_{ev}}_{i=1} (\mathcal{M}_{2})_{i}, \nonumber \\  
\left\langle  2 \right\rangle_{n}      &=&  \dfrac{Q^{\eta_{a}}_{n,1} \left(   Q^{\eta_{b}}_{n,1} \right)^{*} }{\mathcal{M}_{2}} \nonumber \\
\mathcal{M}_{2}&=& S^{\eta_{a}}_{1,1}  S^{\eta_{b}}_{1,1},
\end{eqnarray}
%------------------------------------------------------------------------
where $\langle \langle \, \rangle \rangle$ is the average over particles, {\color{black} then a multiplicity weighted average over events.}

%------------------------------------------------------------------------
\item{The three-particle correlations:}
%C_{n+m,nm}     &=&     \nonumber\\
%------------------------------------------------------------------------
\begin{eqnarray}\label{eq:21-4}
 \left\langle  \left\langle  3 \right\rangle \right\rangle_{knm} &=& \langle\langle  \cos(k\phi^{a}_1 - n\phi^{b}_2 - m\phi^{b}_3 \rangle\rangle, \\  \nonumber
  &=&   \sum^{N_{ev}}_{i=1} (\mathcal{M}_{3})_{i} \left\langle  3 \right\rangle_{{knm}, i} / \sum^{N_{ev}}_{i=1} (\mathcal{M}_{3})_{i}, \nonumber
\end{eqnarray} 
where $k=n+m$,
\begin{eqnarray}\label{eq:21-4x}
\left\langle  3 \right\rangle_{knm}      &=&  \dfrac{ \left(    Q^{\eta_{a}}_{n,1}  Q^{\eta_{a}}_{m,1}  -   Q^{\eta_{a}}_{k,2}      \right)          \left(    Q^{\eta_{b}}_{k,1}   \right)^{*}    }
{ \mathcal{M}_{3} }, \nonumber \\
\mathcal{M}_{3}&=& \left(  S^{\eta_{a}}_{2,1}   - S^{\eta_{a}}_{1,2}    \right)   S^{\eta_{b}}_{1,1},
\end{eqnarray} 
%------------------------------------------------------------------------

%------------------------------------------------------------------------
\item{The four-particle correlations:}
%------------------------------------------------------------------------
%v_{n}^{2} v_{m}^{2}
\small{
\begin{eqnarray}\label{eq:21-5}
\langle \langle  4 \rangle   \rangle_{nm}   &=& \langle\langle  \cos(n\phi^{a}_1 + m\phi^{a}_2 - n\phi^{b}_3 - m\phi^{b}_4 \rangle\rangle, \\  \nonumber
   &=&   \sum^{N_{ev}}_{i=1} (\mathcal{M}_{4})_{i} \left\langle  4 \right\rangle_{{nm}i} / \sum^{N_{ev}}_{i=1} (\mathcal{M}_{4})_{i}, \nonumber\\
\left\langle  4 \right\rangle_{nm}      &=&  \dfrac{ \left(    Q^{\eta_{a}}_{n,1}  Q^{\eta_{a}}_{n,1}  -  S^{\eta_{a}}_{1,2} Q^{\eta_{a}}_{2n,1}      \right)          \left(    Q^{\eta_{b}}_{m,1}  Q^{\eta_{b}}_{m,1}  -  S^{\eta_{b}}_{1,2} Q^{\eta_{b}}_{2m,1}      \right)^{*}    }
{ \mathcal{M}_{4} }, \nonumber \\
\mathcal{M}_{4}&=& \left(  S^{\eta_{a}}_{2,1}   - S^{\eta_{a}}_{1,2}    \right)   \left(  S^{\eta_{b}}_{2,1}   - S^{\eta_{b}}_{1,2}    \right).
\end{eqnarray} 
}
%------------------------------------------------------------------------
%------------------------------------------------------------------------
\end{itemize}

Using the two- and four-particle correlations we can write the four-particle n$^{th}$ flow harmonics, and the normalized symmetric cumulants as:
%
%\sqrt[4]{}
\begin{eqnarray}
\label{eq:c2-1}
v^{4}_n\{4\}  &=&  2\left\langle\left\langle 2 \right\rangle\right\rangle_{n}^2  -  \left\langle\left\langle 4 \right\rangle\right\rangle_{nn} ,
\end{eqnarray}
%------------------------------------------------------------------------
%------------------------------------------------------------------------
%\sqrt[4]{}
\begin{eqnarray}
\label{eq:c23-1}
{\rm NSC}(n,m)  &=&   \frac{ \left\langle\left\langle 4 \right\rangle\right\rangle_{nm} - \left\langle\left\langle 2 \right\rangle\right\rangle_{n}  \left\langle\left\langle 2 \right\rangle\right\rangle_m }{\left\langle\left\langle 2 \right\rangle\right\rangle_{n}  \left\langle\left\langle 2 \right\rangle\right\rangle_{m}}  .
\end{eqnarray}
%------------------------------------------------------------------------

The benefit of using the two-subevents technique is that it assists the reduction of the near-side non-flow correlations resulting from   resonance decays, Bose-Einstein correlations, and the fragments of individual jets~\cite{Magdy:2020bhd}. \\
%------------------------------------------------------------------------

%------------------------------------------------------------------------
%where $\langle \langle \, \rangle \rangle$ is the average over particles and average over events,  and $\varphi_{i}$ is the azimuthal angle of the $\textit{i}^{th}$ particle. 
%------------------------------------------------------------------------

%------------------------------------------------------------------------
\paragraph{Linear and non-linear contributions:\\}
%------------------------------------------------------------------------
The inclusive \fourth which contains the linear and the non-linear contributions is given by the two-particle correlations Eq.~(\ref{eq:21-3}):
\begin{eqnarray}\label{eq:2-4x}
 v^{\rm Inclusive}_{4}   &=&   v_{4}\{2\}.
\end{eqnarray}

The non-linear contribution to \fourth can be given as~\cite{Yan:2015jma,Bhalerao:2013ina}:
%------------------------------------------------------------------------
\begin{eqnarray}\label{eq:2-4}
v_{4}^{\rm Non Linear} &=&  \frac{\left\langle  \left\langle  3 \right\rangle \right\rangle_{422}} {\sqrt{\langle \langle  4 \rangle \rangle_{22}}},\\
                     &\sim & \langle v_{4} \, \cos (4 \Psi_{4} - 2\Psi_{2} - 2\Psi_{2}) \rangle,  \nonumber
\end{eqnarray}
%
%\\ 
% 
%
and the linear contribution to \fourth~\cite{Yan:2015jma,Magdy:2020bhd} can be expressed as:
\begin{eqnarray}\label{eq:2-5x}
v_{4}^{Linear}  &=& \sqrt{ (v^{\rm Inclusive}_{4})^{\,2} - (v^{\rm Non Linear}_{4})^{\,2}  }.
\end{eqnarray}
%
%Equation (\ref{eq:2-5x}) suggests that $\textit{v}_{4}^{\rm Non Linear}$ and $\textit{v}_{4}^{\rm Linear}$ are independent~\cite{Yan:2015jma,Magdy:2020bhd}.

%------------------------------------------------------------------------
The non-linear response coefficient ($\chi_{4,22}$), which quantify the  mode-coupling contributions to the \fourth, is defined as:
\begin{eqnarray}\label{eq:2-7}
\chi_{4,22} &=& \frac{v^{\rm Non Linear}_{4}} {\sqrt{ \langle \langle  4 \rangle \rangle_{22} }}.
\end{eqnarray}
%------------------------------------------------------------------------

%------------------------------------------------------------------------
The correlations between different order flow symmetry planes ($\rho_{4,22}$)~\cite{Acharya:2017zfg}  can be given as:
%\frac{v^{\rm Non Linear}_{4}}{v^{\rm Inclusive}_{4}} 
\begin{eqnarray}\label{eq:2-6}
\rho_{4,22} &=& \frac{\left\langle  \left\langle  3 \right\rangle \right\rangle_{422}} {\sqrt{\langle \langle  4 \rangle \rangle_{22}   \langle \langle  2 \rangle \rangle_{4} }} \sim \langle  \cos (4 \Psi_{4} - 4 \Psi_{2}) \rangle.
\end{eqnarray} \\
%------------------------------------------------------------------------
%------------------------------------------------------------------------

%------------------------------------------------------------------------
\paragraph{The flow transverse-momentum correlations:\\}
%------------------------------------------------------------------------
The \pcct correlation coefficient (Eq.~\ref{eq:1})  contains the $v_n$ and $\langle p_{T} \rangle$ variances and covariances that utilize the two- and multi-particle correlations. 
%Such correlations could be impacted by non-flow effects resulting from, Bose-Einstein correlations, resonance decays, and the fragments of individual jets~\cite{Jia:2013tja}.  These non-flow effects  can be reduced using the subevent cumulant method~\cite{Jia:2017hbm,Huo:2017nms,Zhang:2018lls,Magdy:2020bhd}.
%------------------------------------------------------------------------
%The \pcct correlation coefficient (Eq.~\ref{eq:1})  involves the evaluation of variances and covariances that employ two- and multi-particle correlations. These correlations could be subject to non-flow effects resulting from resonance decays, Bose-Einstein correlations, and the fragments of individual jets~\cite{Jia:2013tja}. However, such non-flow effects are dominated by particles emitted within a localized $\mathrm{\eta}$-region and can be minimized using the subevent cumulant method~\cite{Jia:2017hbm,Huo:2017nms,Zhang:2018lls,Magdy:2020bhd}. The efficacy of the method has been demonstrated for two- and multi-particle correlations~\cite{Jia:2017hbm,Huo:2017nms,Magdy:2020bhd}.
%------------------------------------------------------------------------

%------------------------------------------------------------------------
The $v_{2}^{2}$ variance can be given as:
%------------------------------------------------------------------------
\begin{eqnarray}\label{eq:2-1}
    {\rm Var}(v_{2}^{2}) &\sim& v_{2}\{2\}^{4} - v_{2}\{4\}^{4},
\end{eqnarray}
%------------------------------------------------------------------------
where $v_{2}\{2\}$ and $v_{2}\{4\}$  are the two- and four-particle elliptic flow using the subevent method~\cite{Jia:2017hbm} (see Eqs.~\ref{eq:21-3} and \ref{eq:c2-1}).

%------------------------------------------------------------------------
The variance of the  $\langle p_{T} \rangle$~\cite{Abelev:2014ckr},  evaluated in the range  $|\eta_{B}|<0.35$ {\color{black}using the two particle correlation method~\cite{ALICE:2018jco}},  given as:
%------------------------------------------------------------------------
\small{
\begin{eqnarray}\label{eq:2-5}
    c_k  = \left\langle  \frac{1}{N_{\rm pair}} \sum_{B}\sum_{B^{\prime}\neq B} (p_{T,B} - \langle \langle p_{T} \rangle \rangle )  (p_{T,B^{\prime}} - \langle \langle p_{T} \rangle \rangle) \right\rangle,  \nonumber \\
\end{eqnarray}
}
%------------------------------------------------------------------------
where  $\langle  \rangle$ is an average over all events. {\color{black} The condition $B^{\prime}\neq B$ is used to remove self-correlations.} The event mean $p_T$, is given as,
%------------------------------------------------------------------------
\begin{eqnarray}\label{eq:2-6}
     \langle p_{T} \rangle  =  \sum^{M_{B}}_{i=1} p_{T,i}  /  M_{B},
\end{eqnarray}
where $M_{B}$ is the number of tracks in subevent $B$.

The  covariance   of $v_{2}^{2}$ and the $\langle p_{T} \rangle$ (${\rm cov}(v_{2}^{2},\langle p_{T} \rangle)$) are calculated through the three-subevents method~\cite{Aad:2019fgl,Zhang:2021phk} as,
%------------------------------------------------------------------------
\small{
\begin{eqnarray}\label{eq:2-7}
{\rm cov}(v_{2}^{2},\langle p_{T} \rangle) &=&  {\rm Re} \left( \left< \sum_{A,C} e^{i2(\phi_{A} - \phi_{C})} \left( \langle p_{T} \rangle - \langle \langle p_{T} \rangle \rangle \right)_{B} \right> \right). \nonumber \\
\end{eqnarray}
}
%------------------------------------------------------------------------
The $\rho(v^{2}_{2},\langle p_{T} \rangle)$  coefficient~\cite{Giacalone:2020byk,Lim:2021auv,Bozek:2016yoj,Bozek:2020drh,Schenke:2020uqq,ATLAS:2021kty} can be given using Eqs.~\ref{eq:2-1}, \ref{eq:2-5} and \ref{eq:2-7};
%,,
%------------------------------------------------------------------------
\begin{eqnarray}\label{eq:2-8}
    \rho(v^{2}_{2},\langle p_{T} \rangle) = \frac{{\rm cov}(v_{2}^{2},\langle p_{T} \rangle)}{\sqrt{{\rm Var}(v_{2}^{2})} \sqrt{c_k}}.
\end{eqnarray}
%------------------------------------------------------------------------
%------------------------------------------------------------------------

%\begin{widetext}
%\end{widetext}
%------------------------------------------------------------------------

%------------------------------------------------------------------------

%------------------------------------------------------------------------
\section{Results and discussion}\label{sec:3}
%------------------------------------------------------------------------
%------------------------------------------------------------------------
%are sensitive to initial- and final-state effects.
%------------------------------------------------------------------------
%------------------------------------------------------------------------

%------------------------------------------------------------------------
\begin{figure}[h] 
\begin{center}
\includegraphics[width=1.05 \linewidth, angle=-0,keepaspectratio=true,clip=true]{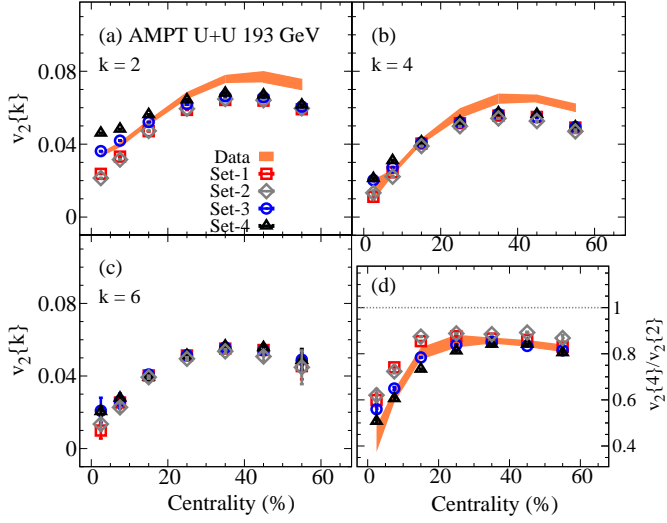}
\vskip -0.3cm
\caption{Centrality dependence of $v_{2}\lbrace 2\rbrace$ (a), $v_{2}\lbrace 4\rbrace$ (b), $v_{2}\lbrace 6\rbrace$ (c), and $v_{2}\lbrace 4\rbrace / v_{2}\lbrace 2\rbrace$  computed with the AMPT model sets Tab.~\ref{tab:1} for U+U collisions at \roots = 193~GeV. The bands represent the experimental data reported in Refs.~\cite{STAR:2019zaf,STAR:2022gki}.
 \label{fig:1} }
\end{center}
\vskip -0.5cm
\end{figure}
%------------------------------------------------------------------------
%------------------------------------------------------------------------
\begin{figure}[h]
\begin{center}
\includegraphics[width=1.05 \linewidth, angle=-0,keepaspectratio=true,clip=true]{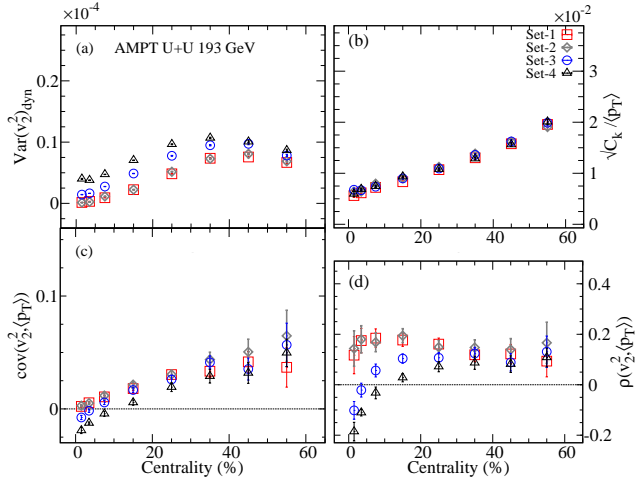}
\vskip -0.3cm
\caption{Comparison of the centrality dependence of ${\rm Var}(v_{n}^{2})_{dyn}$ (a), $\sqrt{c_{k}}/\langle p_{T}\rangle$ (b), ${\rm cov}(v_{n}^{2},\langle p_{T} \rangle)$ (c) and $\rho(v^{2}_{n},\langle p_{T} \rangle)$ (d) computed from the AMPT model sets given in Tab.~\ref{tab:1}, for U+U collisions at \roots = 193~GeV.\label{fig:2}
  }
\end{center}
\vskip -0.5cm
\end{figure}
%------------------------------------------------------------------------
%------------------------------------------------------------------------
\begin{figure}[h] 
\begin{center}
\includegraphics[width=0.80 \linewidth, angle=-0,keepaspectratio=true,clip=true]{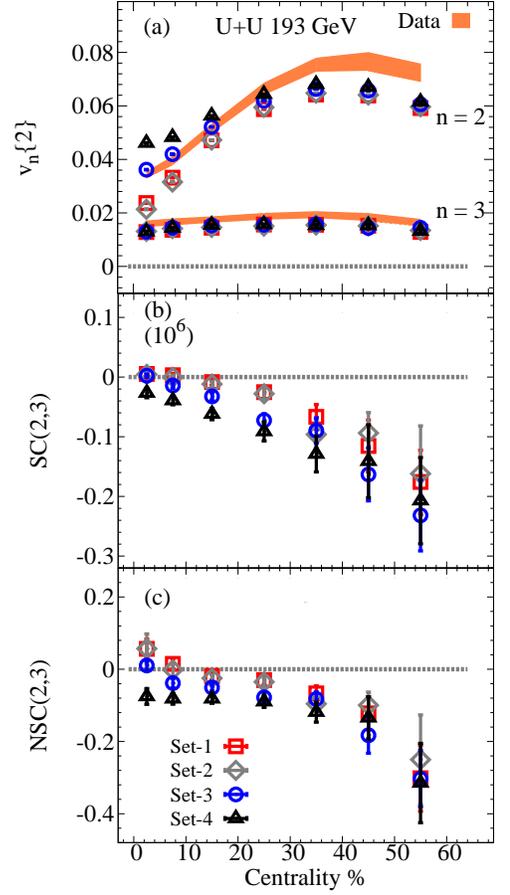}
\vskip -0.3cm
\caption{Centrality dependence of $v_{n}\lbrace 2\rbrace$ panel (a), the ${\rm SC}(2,3)$ panel (b), and the ${\rm NSC}(2,3)$ panel (c)  computed with the AMPT model sets given in Tab.~\ref{tab:1}, for U+U collisions at \roots = 193~GeV. The bands represent the experimental data reported in Refs.~\cite{STAR:2019zaf,STAR:2022gki}.
\label{fig:3}
}
\end{center}
\vskip -0.5cm
\end{figure}
%------------------------------------------------------------------------
%------------------------------------------------------------------------
%------------------------------------------------------------------------
\begin{figure*}[t]
\begin{center}
\includegraphics[width=0.88 \linewidth, angle=-0,keepaspectratio=true,clip=true]{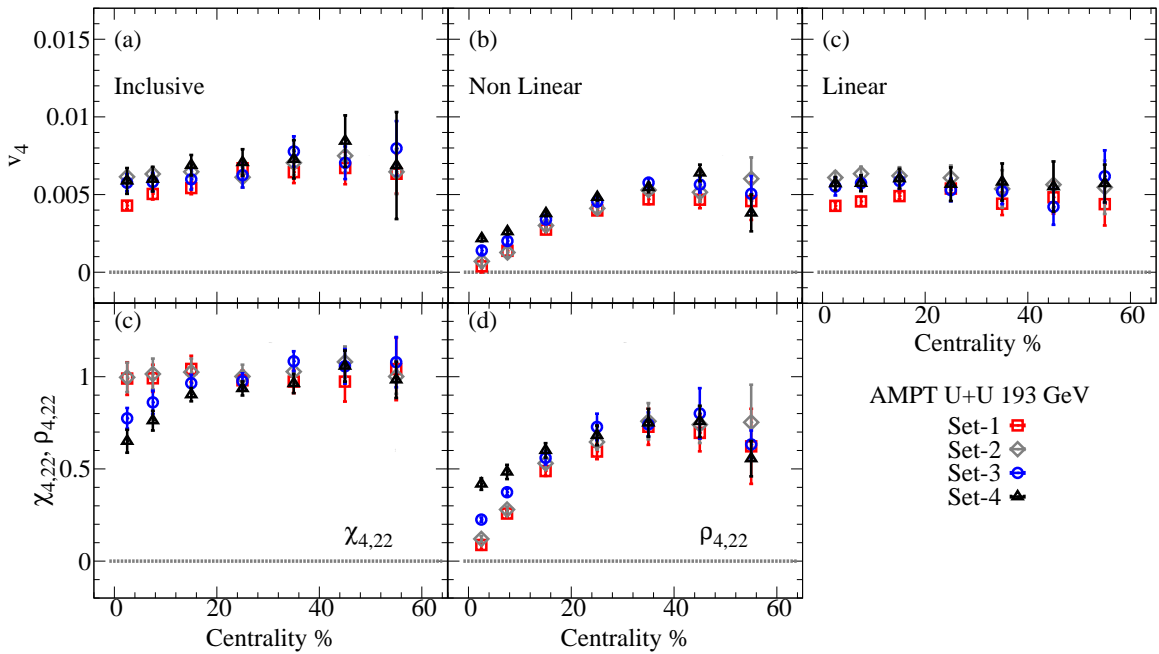}
\vskip -0.93cm
\caption{ Comparison of the centrality-dependent inclusive, linear and non-linear \fourth panels (a)--(c), $\chi_{4,22}$ panel (d) and $\rho_{4,22}$ panel (e) obtained from the AMPT model sets given in Tab.~\ref{tab:1}, for U+U collisions at \roots = 193~GeV. \label{fig:4}
  }
\end{center}
\vskip -0.5cm
\end{figure*}
%------------------------------------------------------------------------

%------------------------------------------------------------------------
Figure~\ref{fig:1} shows a comparison of the centrality dependence of $v_{2}\lbrace 2\rbrace$ (a), $v_{2}\lbrace 4\rbrace$ (b) , $v_{2}\lbrace 6\rbrace$ (c) and the ratios $v_{2}\lbrace 4\rbrace/v_{2}\lbrace 2\rbrace$ (d) obtained with the AMPT model sets shown in Table~\ref{tab:1}.
Panels (a)--(c) show that $v_{2}\lbrace 2\rbrace$ is much more sensitive to deformation than $v_{2}\lbrace 4\rbrace$ and $v_{2}\lbrace 6\rbrace$ and shows a sizable increase with {\color{black} the quadruple deformation given by} $\beta_2$. {\color{black}Such observations reflect the $v_{2}\lbrace 2\rbrace$ and $v_{2}\lbrace 4\rbrace$ dependence on $\beta_{2}^{2}$ and $\beta_{2}^{4}$ respectively as suggested by Refs~\cite{Giacalone:2018apa,Giacalone:2021uhj,Giacalone:2021udy,Jia:2021tzt,Giacalone:2017uqx}.}
 In contrast, $v_{2}\lbrace k\rbrace$ show no sensitivity to the {\color{black}hexadecapole deformation given by} $\beta_4$, {\color{black}confirming that $v_{2}\lbrace k\rbrace$ depends only on $\beta_2$.}
The bands in panels (a), and (b) show the  experimental measurements constructed from Refs.~\cite{STAR:2019zaf,STAR:2022gki}. {\color{black}The AMPT calculations show poor quantitative agreement with the experimental measurements.}

{\color{black}The ratio $v_{2}\lbrace 4\rbrace/v_{2}\lbrace 2\rbrace$ gives the strength of the elliptic flow fluctuations~\cite{Manly:2005zy,Magdy:2020gxf,Rao:2019vgy}. Note that $v_{2}\{4\}/ v_{2}\{2\} \approx 1.0$ suggests small, if any, fluctuations, whereas $v_{2}\{4\}/ v_{2}\{2\} < 1.0$ implies more fluctuations as this ratio decreases. In addition, nuclear deformation increases the initial state fluctuations. Fig.~\ref{fig:1}(d) shows the ratios $v_{2}\lbrace 4\rbrace/v_{2}\lbrace 2\rbrace$; the band shows the experimental measurements constructed from Refs.~\cite{STAR:2019zaf,STAR:2022gki}. 
The results show that the elliptic flow fluctuations increase with the initial-state fluctuations caused by the increase in the $\beta_2$ value.}
%
%indicating that  the combined measurements of $v_{2}\lbrace 2\rbrace$ and $v_{2}\lbrace 4\rbrace/v_{2}\lbrace 2\rbrace$ provides more stringent constraints for $\beta_2$ than the seperate measurements of  $v_{2}\lbrace 2\rbrace$ and $v_{2}\lbrace 4\rbrace/v_{2}$.
%
 %In addition the AMPT sets with the $\beta_2$ $>0.0$ show better agreement with the experimental data.
%
 The results for $v_{2}\lbrace k\rbrace$ are sensitive to initial-state effects (\textit{i.e.}, initial-state eccentricity and initial-state eccentricity fluctuations), and final-state effects (\textit{i.e.}, viscous attenuation). Therefore, they are more suitable for constraining the interplay between final- and initial-state effects. 
By contrast, the elliptic flow fluctuations are sensitive to the initial-state eccentricity and its fluctuations. Therefore, combining the results of the $v_{2}\lbrace 2\rbrace$ and the ratio $v_{2}\lbrace 4\rbrace/v_{2}\lbrace 2\rbrace$ can be used to add a simultaneous constraint on the initial and final-state effects.
%
%Our results show that the flow fluctuations increase with the $\beta_{2}$ values.
%------------------------------------------------------------------------

%------------------------------------------------------------------------
In prior work, the correlation between $v_2$ and the event mean $p_{T}$ ($\rho(v^{2}_{2},\langle p_{T} \rangle)$) has been shown to be sensitive to the nuclear deformation~\cite{Giacalone:2019pca,Giacalone:2020byk,Bozek:2016yoj,Bozek:2020drh,Schenke:2020uqq,Giacalone:2020dln,Giacalone:2020awm,Jia:2021wbq}.
Fig.~\ref{fig:2} compares the $\beta_2$ dependence of ${\rm Var}(v_{2}^{2})$ (a), $\sqrt{c_{k}}/\langle p_{T}\rangle$ (b), ${\rm cov}(v_{2}^{2},\langle p_{T} \rangle)$ (c) and $\rho(v^{2}_{2},\langle p_{T} \rangle)$ (d) respectively for U$+$U collisions at \roots = 193~GeV from the AMPT model.
The results indicate that in central collisions, ${\rm Var}(v_{2}^{2})$ (panel (a)) increases with $\beta_2$ while ${\rm cov}(v_{2}^{2},\langle p_{T} \rangle)$ (panel (c)) and  $\rho(v^{2}_{2},\langle p_{T} \rangle)$ (panel (d)) decrease with  $\beta_2$ and even become negative in more central collisions. {\color{black}Such observations can be explained by considering the core of the ${\rm cov}(v_{2}^{2},\langle p_{T} \rangle)$ dependence on $\beta_2$~\cite{Jia:2021qyu,Bally:2021qys} (i.e., ${\rm cov}(v_{2}^{2},\langle p_{T} \rangle)$ $\sim$ $a_{0}$ - $a_{1}$ $\beta_{2}^{3}$). These results agree qualitatively with the recent STAR experiment preliminary measurements~\cite{Zhang:2022sgk}.}
 The values for $\sqrt{c_{k}}/\langle p_{T}\rangle$ (panel (b)) are relatively insensitive to $\beta_2$, {\color{black}which does not agree with the preliminary measurements of the STAR experiment~\cite{Zhang:2022sgk}.}
  In addition, these results show no sensitivity to the $\beta_4$ variations given by AMPT Set-$2$. %As noted earlier, $\rho(v^{2}_{2},\langle p_{T} \rangle)$ is known to be sensitive to the $p_T$ and $\eta$ cuts as well as the centrality definitions~\cite{ATLAS:2022dov,ALICE:2021gxt}.
%------------------------------------------------------------------------
The ${\rm Var}(v_{2}^{2})$ and ${\rm cov}(v_{2}^{2},\langle p_{T} \rangle)$ results in Fig.~\ref{fig:2} are sensitive to the interplay between final- and initial-state effects. However, the $\rho(v^{2}_{2},\langle p_{T} \rangle)$  {\color{black}
is suggested to leverage the correlation between the eccentricity-driven $v^{2}_{2}$ and the transverse size of the overlap region given by the $\langle p_{T} \rangle$~\cite{Bozek:2012fw}.}
%primarily sensitive to the initial-state correlations.  
Therefore, the combined measurements of $v_{2}\lbrace 2\rbrace$, $v_{2}\lbrace 4\rbrace/v_{2}\lbrace 2\rbrace$ and $\rho(v^{2}_{2},\langle p_{T} \rangle)$ would be expected to provide even more stringent constraints for $\beta_2$ and the interplay between final- and initial-state effects.
%------------------------------------------------------------------------
%------------------------------------------------------------------------

Supplemental constraints for the initial-state deformation can be obtained in tandem via the ${\rm SC}(2,3)$ and the ${\rm NSC}(2,3)$ correlators.
Figure.~\ref{fig:3} shows a comparison of the centrality dependence of  $v_{n}\lbrace 2 \rbrace$ (panel (a)) and ${\rm NSC}(2,3)$ (panel (b)). The $v_{2}\lbrace 2 \rbrace$ values from the AMPT simulations show a clear sensitivity to $\beta_2$ as in Fig.~\ref{fig:1}. By contrast, $v_{3}\lbrace 2 \rbrace$ shows the expected insensitivity to $\beta_2$. Both quantities show no sensitivity to the $\beta_4$ variations given by the AMPT Set-$2$. The bands in panel (a) represent the experimental data~\cite{STAR:2019zaf,STAR:2022gki} that shows similar trend to the AMPT calculations presented.
{\color{black}The ${\rm SC}(2,3)$ (panel (c))  which contains the $\epsilon_{2}$ and the $\epsilon_{3}$ variances and correlations indicate a negative value that increase with $\beta_2$. Similarly, the dimensionless flow harmonic correlations, ${\rm NSC}(2,3)$ (panel (c)) indicate an anti-correlation  between $v_2$ and $v_3$ which grows with $\beta_2$, suggesting that its sensitivity to the initial-state deformation can be employed as a supplemental constraint.}
These ${\rm NSC}(2,3)$ trends, in tandem with the results shown in Figs.~\ref{fig:1} and~\ref{fig:2} could provide more stringent constraints for the influence of nuclear deformation on the initial-state correlations and fluctuations as well as it's interplay with final-state effects.

%------------------------------------------------------------------------
The centrality dependence of the inclusive, linear and non-linear \fourth (panels (a)--(c)) as well as the non-linear response coefficients, $\chi_{4,22}$ (panel (d)), and the correlations of the event plane angles, $\rho_{4,22}$ are shown in Fig.~\ref{fig:4}. {\color{black} The results indicate that, the inclusive \fourth depends on $\beta_4$ and $\beta_2$ ($\left( \fourth^{Inclusive} \right)^2 $ $\sim$ $a3$ $\beta_{4}^{2}$ + $a4$ $\beta_{2}^{4}$) which in line with the results presented in Ref~\cite{Jia:2021tzt}.} In addition, the AMPT calculations of the  linear \fourth shows a sensitivity to the $\beta_4$ variation in central collisions. On other hand,  the non-linear contribution of \fourth, which has the weakest contribution to the inclusive \fourth in central collisions, have a sizable dependence on $\beta_2$. 
Therefore, using the linear \fourth can add a constraints on the $\beta_4$ values and the non-linear \fourth can add a constraints on the $\beta_2$ values.

Figure.~\ref{fig:4} (c) shows that the $\chi_{4,22}$ indicate a weak centrality dependence for the non-deformed U$+$U collisions ($\beta_{2} = 0.0$) and a modest centrality dependence, in central collisions, for deformed U$+$U collisions ($\beta_{2} > 0.0$). {\color{black} Such an observation suggests that $\chi_{4,22}$ depends on initial-state effects, which disagrees with Ref~\cite{Yan:2015jma}.} 
 In addition, the $\rho_{4,22}$, that shows stronger event plane correlations in peripheral collisions, Fig.~\ref{fig:4} (d) indicate sizable dependence on $\beta_2$ in central collisions. {\color{black}  The $\rho_{4,22}$ results suggest that it depends on $\beta_2$ similarly to the non-linear \fourth.}
Also, the results of $\chi_{4,22}$ and $\rho_{4,22}$ show no sensitivity to the $\beta_4$ variations given by AMPT Set-$2$. 
The dimensionless coefficients $\chi_{4,22}$($\rho_{4,22}$)  that show a sizable dependence on $\beta_2$ in central collisions (note the indicated increase(decrease) of $\rho_{4,22}$($\chi_{4,22}$) with $\beta_2$) suggesting their value as supplemental constraints to the nuclear deformation effects in U$+$U collisions.
%------------------------------------------------------------------------
%------------------------------------------------------------------------

The results presented in Figs.~\ref{fig:1}--\ref{fig:4} as well as in prior studies~\cite{Magdy:2021ocp,Magdy:2021cci,Magdy:2021sba} indicate that the correlators studied fall into two broad categories: (I)  correlators that are sensitive to the interplay between initial- and final-state effects ($v_{n}\lbrace k\rbrace$, linear/non-linear $v_{4}$,  ${\rm Var}(v_{n}^{2})_{dyn}$, $\sqrt{c_{k}}$, and ${\rm cov}(v_{n}^{2},\langle p_{T} \rangle)$). These correlators are sensitive to both initial- and final-state effects which  makes them less constraining for pinning down the initial-state deformation. (II) Dimensionless correlators that are much more sensitive to initial-state eccentricity fluctuations ($v_{2}\lbrace 4\rbrace/v_{2}\lbrace 2\rbrace$) and correlations ($\rho(v^{2}_{2},\langle p_{T} \rangle)$, $\chi_{4,22}$ and ${\rm  NSC}(2,3)$) and the initial-state angular correlations ($\rho_{4,22}$). Therefore,  a possible route to constraining the nuclear deformation in U$+$U collisions would be to:
(a) use the correlators from category (II) to constrain the effects of the nuclear deformation on the initial-state eccentricity fluctuations, initial-state eccentricity correlations and initial-state angular correlations then (b) use the correlators from the first category to constrain the final-state effects and its interplay with the initial-state effects.
%
%
%initial-state effects and hence more discerning for the nuclear deformation in the U$+$U collisions.  As noted earlier, $\rho(v^{2}_{2},\langle p_{T} \rangle)$ is known to be sensitive to the $p_T$ and $\eta$ cuts as well as the centrality definitions~\cite{ATLAS:2022dov,ALICE:2021gxt}. The caveats in using only the first set or the  $\rho(v^{2}_{2},\langle p_{T} \rangle)$  to constrain the nuclear deformation can be accounted for via: (I) using a supplemental set of dimensionless quantities as suggested in this work or (II) using an isobaric ratios between two isobars as suggested in Ref~\cite{Zhang:2021kxj}. 

%------------------------------------------------------------------------
%In addition The set of dimensionless quantities reported in the current work are in good qualitative agreement with the STAR collaboration preliminary measurements reported in Refs.~\citep{mytalk,mytalk1,STARtalk}. This suggest that a precise set of measurements for $v_{n}\lbrace 2\rbrace$, $v_{2}\lbrace 2\rbrace$/$v_{2}\lbrace 4\rbrace$, $\rho(v^{2}_{2},\langle p_{T} \rangle)$, $NSC(2,3)$, the non-linear  $\textit{V}_{4}$, $\chi_{4,22}$, and $\rho_{4,22}$  can be used to constrain the nuclear deformation effects via data-model comparisons.
%------------------------------------------------------------------------

%------------------------------------------------------------------------
\section{Conclusion} \label{sec:4} 
%------------------------------------------------------------------------
 In summary, we have made systematic investigations of the effects of nuclear deformation  on quantities that are effected by the interplay between the initial- and final-state effects as well as quantities  that are more sensitive to the  initial-state effects. 
 In the framework of the AMPT model we presented the $\beta_2$ and $\beta_4$ dependence of the  $v_{n}\lbrace k\rbrace$, $\rho(v^{2}_{2},\langle p_{T} \rangle)$, flow fluctuations and correlations, linear and non-linear contributions to the $\textit{V}_{4}$, $\chi_{4,22}$ and $\rho_{4,22}$ in U$+$U collisions at \roots = 193 GeV.
 The model predicts characteristic patterns (mostly in central collisions) for the different presented  coefficients consistent with the nuclear deformation effects given by the $\beta_2$ and $\beta_4$ values. 
 These predictions suggest that a precise set of measurements for $v_{n}\lbrace 2\rbrace$, $v_{2}\lbrace 2\rbrace$/$v_{2}\lbrace 4\rbrace$, $\rho(v^{2}_{2},\langle p_{T} \rangle)$, $NSC(2,3)$, the non-linear  $\textit{V}_{4}$, $\chi_{4,22}$, and $\rho_{4,22}$ together can be used to constrain the nuclear deformation effects in U$+$U collisions via data-model comparisons.
%--------------------------------------------------------------------------------------------------------------------------------------------

%--------------------------------------------------------------------------------------------------------------------------------------------
\section*{Acknowledgments}
The author thanks Roy Lacey, Giuliano Giacalone, Jiangyong Jia, and Guo-Liang Ma for the useful discussions and for pointing out important references.
This research is supported by the US Department of Energy, Office of Nuclear Physics (DOE NP),  under contracts DE-FG02-87ER40331.A008.
%--------------------------------------------------------------------------------------------------------------------------------------------
%\bibliographystyle{aipauth4-1}
\bibliography{ref} 
%--------------------------------------------------------------------------------------------------------------------------------------------
\end{document}